\documentclass[12pt]{iopart}
\usepackage{iopams}

\usepackage[dvips]{graphicx}
\usepackage{subfigure}
\usepackage{amsfonts}
\usepackage{amssymb}
\usepackage{amscd}

\expandafter\let\csname equation*\endcsname\relax
\expandafter\let\csname endequation*\endcsname\relax
\usepackage{amsmath}

\usepackage{enumerate}
\usepackage{epsfig}
\usepackage{subfigure}
\usepackage{bm}
\usepackage{xcolor}
\usepackage{amsthm}
\usepackage{framed}
\usepackage{multirow}
\usepackage{mathrsfs,amssymb}
\usepackage{mathtools}
\usepackage{physics}
\usepackage[most]{tcolorbox}

\newtcolorbox[auto counter]{mybox}[2][]{
	enhanced,
	breakable,
	colback=orange!5!white,
	colframe=orange!75!black,
	fonttitle=\bfseries,
	title=Box \thetcbcounter: #2,#1
}

\begin{document}
\title[]{Measurement-device-independent quantum random number generation over 23 Mbps with imperfect single-photon sources}

\author{You-Qi Nie$^{1,2,3,6}$, Hongyi Zhou$^{4,6}$, Bing Bai$^{1,2}$, Qi Xu$^{1,2}$, Xiongfeng Ma$^{5,*}$, Jun Zhang$^{1,2,3,*}$, and Jian-Wei Pan$^{1,2,3}$}

\address{$^1$ Hefei National Research Center for Physical Sciences at the Microscale and School
of Physical Sciences, University of Science and Technology of China, Hefei 230026, China}
\address{$^2$ CAS Center for Excellence in Quantum Information and Quantum Physics,
University of Science and Technology of China, Hefei 230026, China}
\address{$^3$ Hefei National Laboratory, University of Science and Technology of China,
Hefei 230088, China}
\address{$^4$ Institute of Computing Technology, Chinese Academy of Sciences, 100190, Beijing, China}
\address{$^5$ Center for Quantum Information, Institute for Interdisciplinary Information Sciences,
Tsinghua University, Beijing 100084, China}
\address{$^6$ These authors contributed equally to this work}
\address{$^*$ Authors to whom any correspondence should be addressed}
\ead{xma@tsinghua.edu.cn and zhangjun@ustc.edu.cn}
\vspace{10pt}
\begin{indented}
\item[]January 2024
\end{indented}

\begin{abstract}
Quantum randomness relies heavily on the accurate characterization of the generator implementation, where the device imperfection or inaccurate characterization can lead to incorrect entropy estimation and practical bias, significantly affecting the reliability of the generated randomness.
Measurement-device-independent (MDI) quantum random number generation (QRNG) endeavors to produce certified randomness, utilizing uncharacterized and untrusted measurement devices that are vulnerable to numerous attack schemes targeting measurement loopholes. However, existing implementations have shown insufficient performance thus far.
Here, we propose a high-speed MDI-QRNG scheme based on a robust measurement tomography approach against the imperfection of single-photon sources.
Compared with the conventional approach, the decoy-state method is introduced to obtain more accurate tomography results and a tighter lower bound of randomness.
Finally, by using a high-speed time-bin encoding system, we experimentally demonstrated the scheme and obtained a reliable min-entropy lower bound of $7.37 \times 10^{-2}$ bits per pulse, corresponding to a generation rate over 23 Mbps, which substantially outperforms the existing realizations and makes a record in discrete-variable semi-device-independent QRNGs.

\end{abstract}

\date{\today}

\maketitle

\vspace{2pc}

\section{Introduction}
In the realm of scientific applications, randomness stands as a fundamental resource, finding utility in diverse scenarios, from cryptography to scientific experimentation. Quantum random number generation (QRNG) ~\cite{MaQRNGReview16, RMPQRNGReview16} is a well-explored field, making use of the intrinsic uncertainty of quantum phenomena to produce randomness that complies with the principles of quantum mechanics. In contrast to their classical counterparts generated by deterministic algorithms, quantum random numbers possess the essential property of unpredictability.
The primary challenge in QRNG is randomness quantification, as the inherent randomness often differs from the nominal one, typically measured by the Shannon entropy of the output distribution. In traditional QRNG setups, where the devices are trusted and well-characterized, randomness quantification is achieved through proper modeling. However, the imperfections of realistic devices and inaccurate characterizations can lead to erroneous entropy estimates and introduce bias into the output bits. Device imperfections, as in quantum key distribution systems, can potentially be exploited by adversaries, allowing them to partially predict outcomes. In certain QRNG applications, particularly those in the realm of cryptography, such drawbacks can result in security threats.

In order to address the challenges of device imperfections and potential adversarial attacks, a variety of QRNG protocols have been proposed, including device-independent (DI) QRNGs~\cite{Pironio2010,bierhorst2018experimentally,liu2018device} and semi-device-independent (Semi-DI) QRNGs~\cite{Brunner15, Ma16, Cao2015losstolerant}. The DI-QRNG protocol, which generates certified randomness based on the violation of Bell’s inequality without relying on the trustworthiness of devices, faces experimental challenges due to the requirement for efficiency-loophole-free Bell tests, rendering it relatively inefficient. The highest reported generation rate for DI-QRNG currently stands at 13 kbps~\cite{liu2021device}. On the other hand, Semi-DI QRNGs, which tolerate certain untrusted or uncharacterized devices, achieve significantly higher randomness generation rates. For instance, continuous variable heterodyne-based source-independent (SI) QRNGs, employing compact fiber optical setups and integrated photonic chips, have attained generation rates of 17 Gbps~\cite{Avesani2018source} and 20 Gbps~\cite{Bertapelle2023highspeed}, respectively.
In the domain of discrete-variable QRNGs, implementations based on single-photon detection have achieved rates up to 16.5 Mbps~\cite{PhysRevApplied.7.054018}. Measurement-device-independent (MDI) QRNGs, designed to generate certified randomness with untrusted measurement devices, represent a significant advance in addressing measurement loopholes. Several MDI-QRNG schemes have been proposed and demonstrated \cite{Cao2015losstolerant, PhysRevA.95.042340, PhysRevA.95.062305}, with implementations varying from time-bin phase encoding \cite{Nie16} to orbital angular encoding \cite{10.1116/5.0074253}.

One of the major obstacles restricting the performance of randomness generation rate of an MDI-QRNG comes from the source imperfection. In an optical system, a practical source inevitably contains the multi-photon component, which can be exploited by the adversary to extract side information \cite{lutkenhaus2000security}. As a result, one can only make a pessimistic estimation of the secure single photon component based on photon number statistics of the source, which leads to a low randomness generation rate.
To address this issue, we adopt the decoy-state method~\cite{Hwang2003Decoy,lo2005decoy, Wangxb2005decoy}, widely applied in quantum cryptography~\cite{RevModPhys.74.145, RevModPhys.92.025002}.
The key idea of the decoy-state method is to prepare probes randomly with different intensities.
Depending on intensities, these probes are classified into signal states (usually with stronger intensity) and decoy states, which in principle, cannot be distinguished by a single measurement.
The decoy-state method enables us to make a more accurate estimation of the single photon component, and can substantially improve the randomness generation rate.

In this work, we propose and experimentally realize a high-speed MDI-QRNG protocol that utilizes a robust quantum measurement tomography approach in the presence of potential adversaries.
The fundamental concept behind our scheme involves preparing specific quantum states as probes for performing measurement tomography.
Different from conventional tomography procedures, our framework aims to eliminate the possibility of tampering with the tomography results.
A key challenge in increasing the speed of the protocol is ensuring the accuracy of measurement tomography, i.e., randomness quantification, which is severely degraded by the issue of imperfect probes caused by multi-photon components.
To address this issue, we apply the decoy-state method and yield a tighter bound on the single photon component, leading to more accurate tomography results and a tighter lower bound on randomness. By using a high-speed time-bin encoding system, we experimentally achieve a generation rate as high as 23 Mbps, which is a record in discrete-variable semi-DI QRNGs.

\section{Theory}
\subsection{Measurement tomography with imperfect probes}

Our quantum random number generation scheme is based on the measurement tomography approach \cite{lundeen2009tomography,fiuravsek2001maximum, Bisio2009OptimalQT}, which reconstructs POVM elements of the measurement device.
A set of POVM consists of positive semi-definite Hermitian operators $\Lambda_b \succeq 0$.
In the measurement tomography, one needs to prepare a series of test states $\{\rho_j\}_j$ that are called probes.
By collecting the experimental statistics,
\begin{equation}
p(b|j) = \mathrm{tr}(\rho_j \Lambda_b),
\end{equation}
the explicit form of the POVM elements can be reconstructed.

The conventional measurement tomography method requires high-fidelity state preparation \cite{lundeen2009tomography,fiuravsek2001maximum}. Otherwise, the tomography results could be inaccurate.
In practical implementations, such requirements usually cannot be satisfied due to device imperfections.
In an optical platform, where ideal single-photon sources are required to perform measurement tomography, inaccurate calibration arises from the presence of multi-photon components of an imperfect source.
A typical example is the coherent state source with a complex amplitude $\alpha$,
\begin{equation}\label{eq:coherentstate}
\ket{\alpha} = e^{-\frac{|\alpha|^2}{2}}\sum_{n=0}^\infty \frac{\alpha^n}{\sqrt{n!}} \ket{n},
\end{equation}
in the Fock basis $\{\ket{n}\}_n$, which is often used as an approximation of single-photon sources after a phase randomization process \cite{gottesman2004security}. Another example is the spontaneous parametric down-conversion source \cite{burnham1970observation,gisin2002quantum}.

For simplicity, we take the two-outcome qubit POVM as an example.
The qubit POVM tomography can be straightforwardly generalized to high-dimensional ones by applying multi-photon probes.
The general form of the elements of a qubit POVM is given by
\begin{equation}\label{eq:qubitPOVM}
\begin{split}
\Lambda_0^{\mathrm{tom}} &= a_0(I+\vec{n}_0 \cdot \vec{\sigma}) \\
\Lambda_1^{\mathrm{tom}} &= a_1(I+\vec{n}_1 \cdot \vec{\sigma}), \\
\end{split}
\end{equation}
where the coefficients $a_0$ and $a_1$ are real numbers, and $\vec{n}_0$ and $\vec{n}_1=(n_x,n_y,n_z)$ are real vectors satisfying
\begin{equation}\label{eq:constraint}
\begin{split}
a_0,a_1&>0 \\
a_0 + a_1 &= 1 \\
|\vec{n}_0|,|\vec{n}_1| &\leq 1 \\
a_0\vec{n}_0 +a_1\vec{n}_1 & = 0,
\end{split}
\end{equation}
which means the qubit POVM can be identified by four parameters $a_1$, $n_x$, $n_y$ and $n_z$.
We can make a full tomography with a complete probe set $\{\rho_j\}_j$. One possible example of this set is $\rho_1 = \ketbra{0}$, $\rho_2=\ketbra{1}$, $\rho_3=\ketbra{+}$ and $\rho_4=\ketbra{+i}$. Then by calculating the conditional probabilities of output $b$ given the probe $\rho_j$, $p(b|j)$ $(b\in\{0,1\},j\in\{1,2,3,4\})$, we can accurately calculate the parameters $a_1$, $n_x$, $n_y$ and $n_z$ and reconstruct $\{\Lambda^{\mathrm{tom}}_0,\Lambda^{\mathrm{tom}}_1\}$.

In practical implementations, measurement tomography suffers from the imperfection of state preparation.
A phase-randomized weak coherent state is often used as a practical single-photon source. After phase randomization, the source becomes a mixture of Fock states \cite{lo2005decoy},
\begin{equation}\label{eq:Poisson}
\begin{split}
P(\ket{\alpha})&=\frac{1}{2\pi}\int_{0}^{2\pi}\ketbra{\alpha e^{i\phi}}d\phi \\
&= \sum_{n=0}^\infty \frac{e^{-|\alpha|^2}|\alpha|^{2n}}{n!}\ket{n}\bra{n},
\end{split}
\end{equation}
where $|\alpha|^2$ is the mean photon number, and the probability of Fock states $\ket{n}$ follows the Poisson distribution.
Here we use $P(\cdot)$ to denote the phase randomization process.
When $\mu$ is small, the vacuum and single-photon components dominate.
Due to the multi-photon components, we cannot make a full tomography even with a complete probe set. Instead, we can bound the parameters $a_1$, $n_x$, $n_y$ and $n_z$ by the following inequalities,
\begin{equation}\label{eq:bounds_para}
 p_1^L(b|j)\leq \mathrm{tr}(\rho_j \Lambda_b^\mathrm{tom}) \leq p_1^U(b|j),
\end{equation}
where $p_1^L(n|j)$ and $p_1^U(n|j)$ are estimated by the decoy-state method. We consider the simplest case with only one decoy state and denote the mean photon number of signal and decoy states as $\mu$ and $\nu$, respectively.
Recalling Eq.~\eqref{eq:Poisson},
the probability of outputting `1' given a signal state or a decoy-state is given by a linear combination of $n$-photon contributions
\begin{equation}\label{eq:pnu}
\begin{aligned}
p_\mu(1|j)&=e^{-\mu}\sum_{n=0} \frac{\mu^n}{n!} p_n(1|j) \\
p_\nu(1|j)&=e^{-\nu}\sum_{n=0} \frac{\nu^n}{n!} p_n(1|j),
\end{aligned}
\end{equation}
where $p_n(1|j)$ is the conditional probability of outputting ‘1’ given a $n$-photon Fock state. Specifically, $p_1(1|j)$ is the conditional probability given an ideal single-photon source, corresponding to the tomography results with ideal probes, and $p_0(1|j)=p_d$ is the dark count rate.
Similar to the decoy state analysis in quantum cryptography \cite{Ma2005DecoyQKD}, the lower and upper bounds of $p_1(1|j)$ can be estimated by some algebra
\begin{equation}\label{eq:lowerbound}
\begin{aligned}
 \frac{\mu}{\mu\nu-\nu^2}\left(p_\nu(1|j)e^\nu-p_\mu(1|j)e^\mu \frac{\nu^2}{\mu^2}-\frac{\mu^2-\nu^2}{\mu^2}p_d\right)  \le p_1(1|j)
  \le \frac{p_\nu(1|j)}{\nu e^{-\nu}}.
 \end{aligned}
\end{equation}

To address statistical fluctuations in our experiment, we focus on the directly observed quantities, denoted as $M_{\mu,j}$ and $M_{\nu,j}$, which represent the number of '1' outputs for different probes. We apply the Chernoff bound method \cite{zhang2017improved}, suitable for independent and identically distributed (i.i.d.) variables. This method enables the estimation of the upper and lower bounds of expectation values based on their observed values,
\begin{equation}\label{eq:detectionfluc}
\begin{aligned}
&\mathbb{E}^L[{M_{\mu(\nu),j}}]=\frac{M_{\mu(\nu),j}}{1+\delta(M_{\mu(\nu)})} \\
&\mathbb{E}^U[{M_{\mu(\nu),j}}]=\frac{M_{\mu(\nu),j}}{1-\delta(M_{\mu(\nu)})} \\
&\delta(x) = \frac{-3\ln\frac{\epsilon}{2}+\sqrt{-8x\ln\frac{\epsilon}{2}+(\ln\frac{\epsilon}{2})^2}}{2(x+\ln(\frac{\epsilon}{2}))},
\end{aligned}
\end{equation}
where we set the a failure probability to $\epsilon=10^{-10}$ for each use of the Chernoff bound. The upper and lower bounds of $p_\mu(1|j)$ and $p_\nu(1|j)$ are calculated as
\begin{equation}\label{eq:sf}
\begin{aligned}
  p^{L}_{\mu}(1|j)&=\frac{\mathbb{E}^{L}({M_{\mu,j}})}{N\eta_j p_s} \\
  p^{U}_{\mu}(1|j)&=\frac{\mathbb{E}^{U}({M_{\mu,j}})}{N\eta_j p_s} \\
  p^{L}_{\nu}(1|j)&=\frac{\mathbb{E}^{L}({M_{\mu,j}})}{N\eta_j (1-p_s)} \\
  p^{U}_{\nu}(1|j)&=\frac{\mathbb{E}^{U}({M_{\mu,j}})}{N\eta_j (1-p_s)},
\end{aligned}
\end{equation}
where $N$ is the number of total test rounds, $\eta_j$ is the proportion of preparing each probe, satisfying $\sum_j \eta_j=1$, $p_s$ is the probability of choosing signal state, and $1-p_s$ is the probability of choosing decoy state.
Combining Eq.~\eqref{eq:lowerbound} and Eq.~\eqref{eq:sf}, the lower and upper bounds, after considering the statistical fluctuations, are
\begin{equation}\label{eq:lowerandupperbound}
\begin{aligned}
&p^L_{1}(1|j)=\frac{\mu}{\mu\nu-\nu^2}\left(p^L_{\nu}(1|j)e^\nu-p^U_{\mu}(1|j)e^\mu\frac{\nu^2}{\mu^2}-\frac{\mu^2-\nu^2}{2\mu^2}p_d\right)  \\
&p^U_{1}(1|j)=\frac{p^U_{\nu}(1|j)}{\nu e^{-\nu}}.
\end{aligned}
\end{equation}
The accuracy of tomography results can be quantified by the fidelity between two POVMs, $\{\Lambda^{\mathrm{tom}}_n\}_n$ and $\{\Lambda^{\mathrm{sim}}_n\}_n$ \cite{lundeen2009tomography}
\begin{equation}\label{eq:POVMfidelity}
\begin{split}
&  F(\{\Lambda_n^{\mathrm{tom}}\}_n, \{\Lambda_n^{\mathrm{sim}}\}_n) = \min_n  \left( \mathrm{tr}\sqrt{\sqrt{\Lambda_n^{\mathrm{sim}}} \Lambda_n^{\mathrm{tom}}\sqrt{\Lambda_n^{\mathrm{sim}}}}\right)^2.
\end{split}
\end{equation}
Here $\{\Lambda^{\mathrm{tom}}_n\}_n$ is parameterized by $a_1$, $n_x$, $n_y$ and $n_z$ according to Eq.~\eqref{eq:qubitPOVM}, and $\{\Lambda^{\mathrm{sim}}\}_n$ is given by a theoretical model of the measurement settings. We assume that the realistic measurement device is characterized by $\{\Lambda^{\mathrm{sim}}_n\}_n$. We note that both $\Lambda_n^{\mathrm{sim}}$ and $\Lambda_n^{\mathrm{tom}}$ should be normalized so that Eq.~\eqref{eq:POVMfidelity} is well defined.
Then, we consider the lower bound of the fidelity that minimizes Eq.~\eqref{eq:POVMfidelity} over all possible parameters $a_1$, $n_x$, $n_y$ and $n_z$, which can be formulated by the following optimization problem,
\begin{equation}\label{eq:fidelity}
\begin{split}
& \min_{a_1,n_x,n_y,n_z} F(\{\Lambda_n^{\mathrm{tom}}\}_n, \{\Lambda_n^{\mathrm{sim}}\}_n) \\
\mathrm{s.t.} \quad & p_1^L(n|j)\leq \mathrm{tr}(\rho_j \Lambda_n^\mathrm{tom}) \leq p_1^U(n|j), \quad n\in\{0,1\}.
\end{split}
\end{equation}

Here are some remarks. First, our method also holds for distributions other than the Poisson distributions. Second,
the density operator of the imperfect probes, $\rho_j^{\mathrm{real}}$ ($j\in\{1,2,3,4\}$), belongs to an infinite-dimensional Hilbert space, which cannot directly act on the two-dimensional POVM $\{\Lambda^{\mathrm{tom}}_n\}_n$. Instead, we make an estimation of the single-photon component with the experimental data. This enables us to obtain bounds of $\mathrm{tr}(\rho_j \Lambda_n^\mathrm{tom})$, which is similar to the idea of single-photon simulator \cite{yuan2016simulating}. Later, we will give the explicit form of $\rho_j^{\mathrm{real}}$ for our time-bin phase encoding implementation.

\subsection{MDI-QRNG}
The MDI-QRNG protocol with the decoy-state method is summarized in Box~\ref{box:MT}.
The randomness is given by the following optimization problem
\begin{equation}\label{eq:minentropy}
\begin{split}
& \min_{a_1,n_x,n_y,n_z} -\frac{2a_1 \mu}{e^{-\mu}}\log_2 \frac{1+\sqrt{1-n_y^2 - n_z^2}}{2} \\
\mathrm{s.t.} \quad & p_1^L(b|j)\leq \mathrm{tr}(\rho_j \Lambda_b^\mathrm{tom}) \leq p_1^U(b|j),\quad b\in\{0,1\}.
\end{split}
\end{equation}
where the bounds $p_1^L(n|j)$ and $p_1^U(n|j)$ are calculated by the decoy-state method as mentioned in the last section while the target function is from  \cite{Cao2015losstolerant}. As a relaxation of the optimization problem Eq.~\eqref{eq:minentropy}, it is safe to only consider the constraint $p_1^L(1|j)\leq \mathrm{tr}(\rho_j \Lambda_1^\mathrm{tom}) \leq p_1^U(1|j)$. To maximize the generation rate of MDI-QRNG, intensities of the signal and decoy states, i.e., $\mu$ and $\nu$ are optimized.

\begin{mybox}[label={box:MT}]{{The MDI-QRNG protocol.}}
\centering
\begin{enumerate}
\item
Alice randomly prepares a probe $\rho^{\mathrm{real}}_j$ with a random intensity $\mu$ or $\nu$. Then
she sends it to the measurement device.
\item
The measurement device returns $0$, $1$, or an inconclusive result (loss or double click) to Alice.
\item
They repeat steps 1 and 2 for $N$ rounds.
\item
Alice records the inconclusive result as $0$ and calculates the conditional probabilities $p(b|j)$, where $b\in\{0,1\}$ is Alice's record.
\item
Alice calculates the min-entropy lower bound by Eq.~\eqref{eq:minentropy}. 
\end{enumerate}
\end{mybox}

In Eq.~\eqref{eq:lowerandupperbound}, the bounds $p^L_{1}(1|j)$ and $p^U_{1}(1|j)$ depend on $\mu$ and $\nu$. Therefore, we can simplify the notations as $p^L_j(\mu,\nu)$ and $p^U_j(\nu)$, respectively. To calculate Eq.~\eqref{eq:minentropy}, we need the lower bound of $a_1$ denoted as $a_1^L$, and the lower bound of $n_y^2+n_z^2$ denoted as $(n_y^2 + n_z^2)^L$. According to Eq.~\eqref{eq:bounds_para}, we have the following bounds,
\begin{equation}
\begin{aligned}
a_1^L(\mu,\nu) &= \frac{1}{2}\left(p_1^L(\mu,\nu)+p_2^L(\mu,\nu)\right) \\
a_1^U(\mu,\nu) & = \frac{1}{2}\left(p_1^U(\mu,\nu)+p_2^U(\mu,\nu)\right) \\
n_y^L(\mu,\nu) & = \frac{p_4^L(\mu,\nu)}{a_1^U(\mu,\nu)}-1 \\
n_y^U(\mu,\nu) & = \frac{p_4^U(\mu,\nu)}{a_1^L(\mu,\nu)}-1 \\
n_z^L(\mu,\nu) & = \frac{p_1^L(\mu,\nu)-p_2^U(\mu,\nu)}{2a_1^U} \\
n_z^U(\mu,\nu) & = \frac{p_1^U(\mu,\nu)-p_2^L(\mu,\nu)}{2a_1^L}. \\
\end{aligned}
\end{equation}
Then, we only need to solve the following optimization problem,
\begin{equation}\label{eq:opt_min_entropy}
\begin{aligned}
& \max_{\mu,\nu} -\frac{2a_1^L(\mu,\nu) \mu}{e^{-\mu}}\log_2 \frac{1+\sqrt{1-(n_y^2 + n_z^2)^L}}{2} \\
& \mathrm{s.t.} \quad  0\leq \nu <\mu \leq 1,
\end{aligned}
\end{equation}
where $(n_y^2 + n_z^2)^L$ is given by
\begin{equation}
\begin{aligned}
& \min_{\mu,\nu}  n_y^2 + n_z^2\\
\mathrm{s.t.} \quad & n_y^L(\mu,\nu) \leq n_y \leq n_y^U(\mu,\nu) \\
& n_z^L(\mu,\nu) \leq n_z\leq n_z^U(\mu,\nu),
\end{aligned}
\end{equation}
which enables us to determine the optimal intensities for both signal and decoy states.
Here we employ a brute-force search to optimize $\mu$ and $\nu$ in Eq.~\eqref{eq:opt_min_entropy}.
It is important to note that the finite accuracy of the numerical calculation may result in $\mu$ and $\nu$ not being the exact optimal values.
However, this deviation only impacts the performance, and the min-entropy remains accurately estimated without overestimation.
\subsection{Simulation}
In our simulation, we make the following estimations of $M_{\mu,j}$ and $M_{\nu,j}$,
\begin{equation}\label{eq:detectionnumber}
\begin{aligned}
M_{\mu,j}&=N\eta_jp_s p_\mu(1|j) \\
M_{\nu,j}&=N\eta_j(1-p_s)p_\nu(1|j),
\end{aligned}
\end{equation}
where the conditional probabilities are estimated by the following equations in terms of detection efficiency $\eta$ and dark count rate $p_d$,
\begin{equation}\label{eq:simu}
\begin{aligned}
p_{\mu}(1\big| \ket{0}\bra{0})&=p_d \\
p_{\mu}(1\big| \ket{1}\bra{1})&=1-(1-p_d)e^{-\eta \mu} \\
p_{\mu}(1\big| \ket{+}\bra{+})&=1-(1-p_d)e^{-\eta \mu/2} \\
p_{\mu}(1\big| \ket{+i}\bra{+i})&=1-(1-p_d)e^{-\eta \mu/2} \\
p_{\nu}(1\big| \ket{0}\bra{0})&=p_d \\
p_{\nu}(1\big| \ket{1}\bra{1})&=1-(1-p_d)e^{-\eta \nu} \\
p_{\nu}(1\big| \ket{+}\bra{+})&=1-(1-p_d)e^{-\eta \nu/2} \\
p_{\nu}(1\big| \ket{+i}\bra{+i})&=1-(1-p_d)e^{-\eta \nu/2}.
\end{aligned}
\end{equation}

\begin{figure}[htbp]
\centering
\includegraphics[width=8.5 cm]{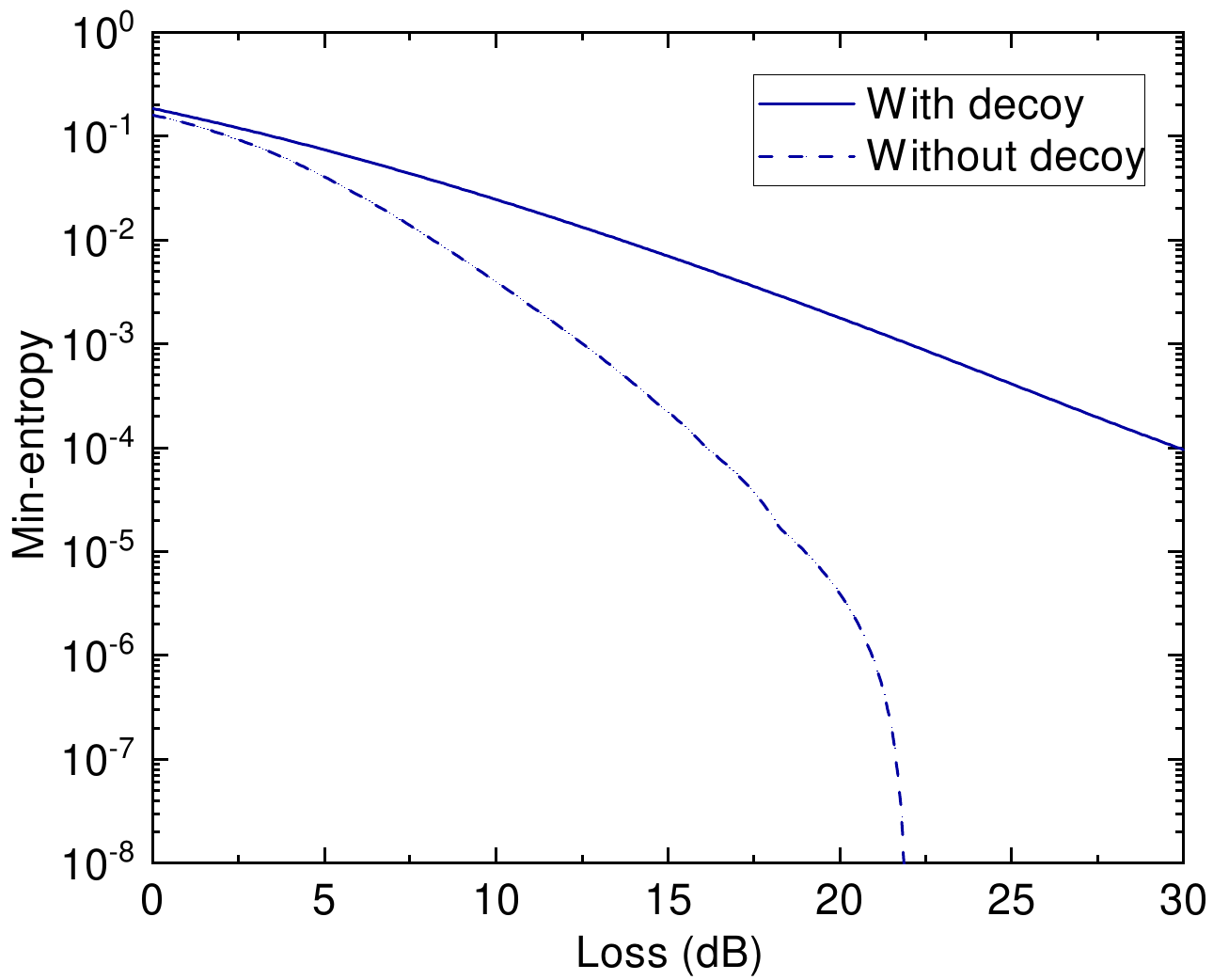}
\caption{Simulation of the min-entropy with and without the decoy-state method.
The simulations ignore the state preparation error and the afterpulse effect of SPAD.}
\label{fig1}%
\end{figure}

The calculation of the randomness lower bound, both with and without the decoy-state method, is conducted by solving the optimization problem presented in Eq.~\eqref{eq:minentropy}. The results of this calculation, depicted in Fig.~\ref{fig1}, illustrate the simulated min-entropy as a function of losses. It is evident from these results that the introduction of the decoy-state method leads to a substantial improvement in the min-entropy, especially in scenarios characterized by high losses. Furthermore, this improvement translates into a significantly enhanced tolerance of the min-entropy against loss, highlighting the effectiveness of the decoy-state method in optimizing quantum randomness generation.

\section{Experiment}
We experimentally demonstrate the high-speed MDI-QRNG protocol by using a 312.5 MHz time-bin encoding optical system.
The target measurement to be characterized by the tomography protocol is $Z$ basis measurement, which consists of a sine wave gating InGaAs/InP single-photon avalanche photodiode (SPAD) with a frequency of 1.25 GHz \cite{Fang2020} and a time-to-digital converter.
The probe states are prepared as phase-randomized time-bin encoding coherent states $\rho_1^{\mathrm{real}} = P(\ket{\alpha}_1\ket{0}_2)$, $\rho_2^{\mathrm{real}} = P(\ket{0}_1\ket{\alpha}_2)$, $\rho_3^{\mathrm{real}}=P(\ket{\alpha/\sqrt{2}}_1\ket{\alpha/\sqrt{2}}_2)$, and $\rho_4^{\mathrm{real}} =P(\ket{\alpha/\sqrt{2}}_1\ket{i\alpha/\sqrt{2}}_2)$, where the subscripts stand for the labels of time bins and $P(\cdot)$ means global phase randomization introduced in Eq.~\eqref{eq:Poisson} that keeps the relative phase between the two modes invariant.
For simplicity, we refer to these imperfect probes $\rho_j^{\mathrm{real}}$ $(j=1,2,3,4)$ as their ideal counterparts $\ket{0}$, $\ket{1}$, $\ket{+}$, and $\ket{+i}$, respectively.

\subsection{Experimental setup}
\begin{figure*}[htp]
\centering
\includegraphics[width=15.5 cm]{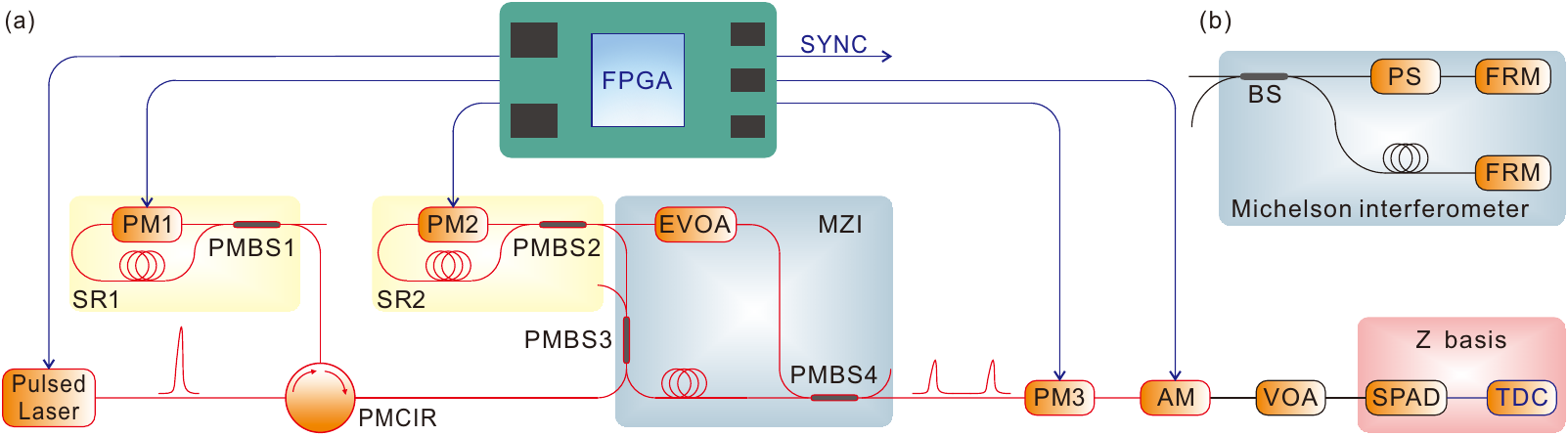}
\caption{(a) Experimental setup for the high-speed MDI-QRNG protocol, including the source and measurement parts. (b) $X$($Y$) basis measurement with a matched Michelson interferometer.
FPGA: field-programmable gate array, SYNC: synchronized signal, PMCIR: polarization-maintaining circulator, PMBS: polarization-maintaining beam splitter, SR: Sagnac ring, PM: phase modulator, EVOA: electrical variable optical attenuator, MZI: Mach-Zehnder interferometer, AM: amplitude modulator, VOA: variable optical attenuator, SPAD: single-photon avalanche diode, TDC: time-to-digital converter, BS: beam splitter, PS: phase shifter, FRM: Faraday rotator mirror.}
\label{fig2}
\end{figure*}

Figure~\ref{fig2}(a) illustrates the experimental setup. A 1550 nm laser diode, driven by narrow pulses, generates phase-randomized laser pulses. These pulses then enter the first Sagnac ring for intensity modulation via a polarization-maintaining circulator \cite{Roberts:18}. The Sagnac ring comprises a polarization-maintaining beam splitter, a phase modulator, and an optical fiber. The phase modulator controls the split ratio between the two output ports of the Sagnac ring. Subsequently, only one of these output ports is connected to the circulator for additional modulation.

The laser pulses then enter the second Sagnac ring and an unbalanced Mach-Zehnder interferometer (MZI) with a time delay of $\thicksim1.6$ ns, forming two time-bin pulses. When the phase modulator in the second Sagnac ring is set to $0$ or $\pi$, the pulse is emitted from one output port, i.e., only an early or late pulse is produced. Then, we realize the preparation of states $\ket{0}$ or $\ket{1}$. When the phase is set to $\pi/2$, the pulse is emitted from two outputs with equal probability. Then, we prepare the state of $\ket{+}$. To prepare quantum states on a $Y$ basis, the third phase modulator is set to $\pi/2$ so that $\ket{+}$ is further modulated into $\ket{+i}$. An additional amplitude modulator is used to increase the visibility of the quantum states $\ket{0}$ or $\ket{1}$. With such a configuration, all the required probe quantum states can be prepared in real-time with different intensities $\mu$ and $\nu$.

In the experiment, the probes $\ket{+}$ and $\ket{+i}$ are indistinguishable when measured in the $Z$ basis. To effectively distinguish between $\ket{+}$ and $\ket{+i}$, measurements in either the $X$ or $Y$ basis are necessary. As depicted in Fig.~\ref{fig2}(b), this is achieved by incorporating a Michelson interferometer before the SPAD, featuring an identical delay as in the MZI. This setup allows for the modification of the $Z$ basis measurement to either an $X$ or $Y$ basis measurement. It is important to note that only the $Z$ basis measurement is required in the implementation of the MDI-QRNG implementation. The $X$ or $Y$ basis is utilized solely to distinguish between the quantum states $\ket{+}$ and $\ket{+i}$ prior to initiating the MDI-QRNG experiment. Therefore, the switch between the $X$, $Y$, and $Z$ bases is performed manually.

\subsection{Measurement tomography result}
\begin{figure*}[tbp]
	\centering
	\includegraphics[width=15 cm]{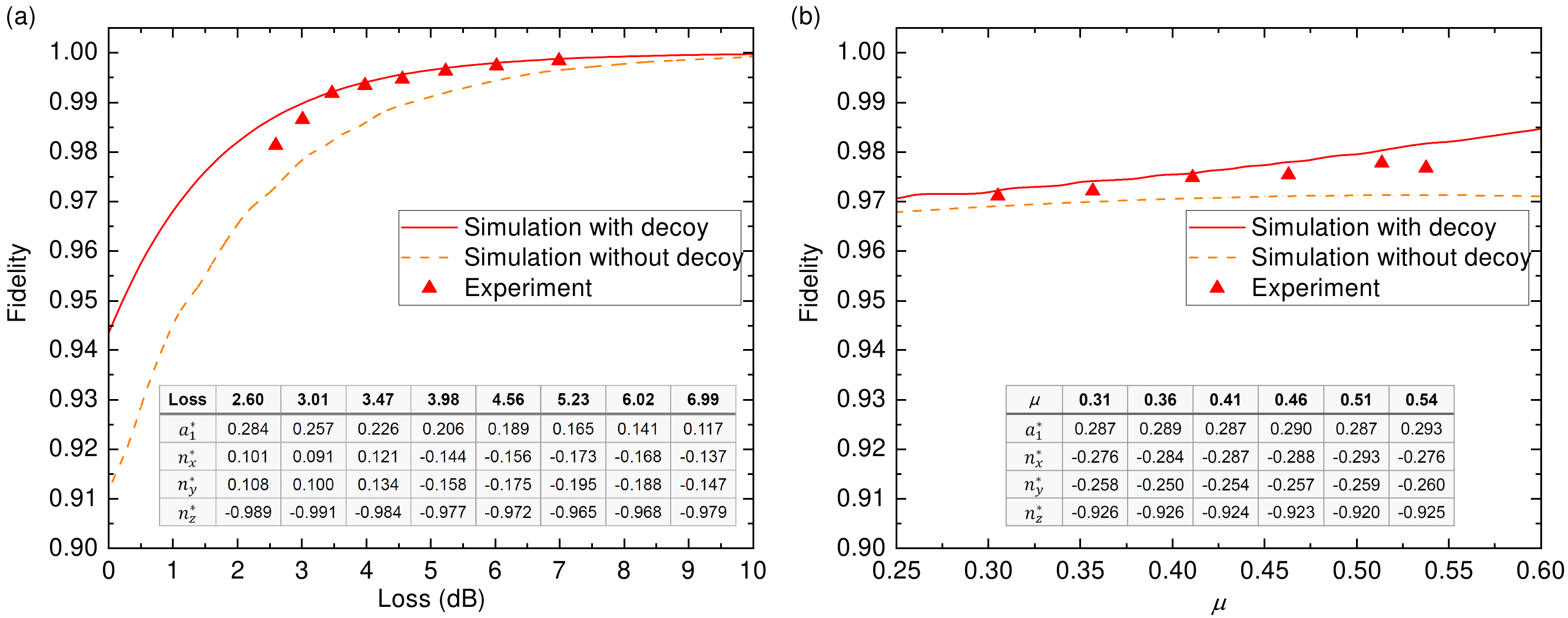}
	\caption{Comparison between simulation results with (solid line) and without (dashed line) the decoy-state method. The triangle symbols represent the experimental results with the decoy-state method. (a) The plot of fidelity result as a function of channel loss with optimized values of $\mu$ and $\nu$. (b) The plot of fidelity result as a function of $\mu$ with an optimized value of $\nu$ given a fixed loss of $2.6$ dB. The corresponding tomography results of $a_1^*$, $n_x^*$, $n_y^*$, and $n_z^*$ are shown in the built-in tables. The simulation parameters include system frequency $f = 312.5$ MHz, dark count probability per gate $p_d = 8\times10^{-5}$, the number of probes $N = 5.625\times10^8$, $\eta_1 = \eta_2 = \eta_3 = \eta_4 = 1/4$, $p_s = 0.5$, and the failure probability is set to $\epsilon=10^{-10}$ for each use of the Chernoff bound. Here, we considered the afterpulse effect, whose probability varies with the detection loss. The error rate caused by the afterpulse effect is set to 1\% in (a) according to average afterpulse probability and 3\% in (b) according to the afterpulse probability at 2.6 dB detection loss, corresponding to additional 1\% and 3\% counts of outputting ‘1’ given different probes, respectively.}
	\label{fig3}%
\end{figure*}

To perform the measurement tomography, four time-bin quantum states of $\ket{0}$, $\ket{1}$, $\ket{+}$, and $\ket{+i}$ with intensities of $\mu$ or $\nu$ are prepared as probes and sent to the device for $Z$ basis measurement.
We perform simulations by comparing the two cases with and without the decoy-state method by solving Eq.~\eqref{eq:fidelity}, as shown in Fig.~\ref{fig3}.
Furthermore, we experimentally verify the protocol's performance with different losses and probe intensities, respectively. It turns out that our experiment results achieve high fidelity, which implies an accurate tomography.

\subsection{MDI-QRNG result}
In the implementation of MDI-QRNG, the setup is operated either in a generation mode or in a tomography mode. In the generation mode, a fixed quantum state $\ket{+}$ with an intensity of $\mu$ is prepared for randomness generation. In the tomography mode, the measurement tomography protocol is performed. In both modes, the untrusted measurement part is set to $Z$ basis for ease of implementation and high detection efficiency. Thus there is no measurement switching. However, the required probe states are randomly prepared in real-time. The proportion of the tomography mode is reduced to a considerably low level to increase the randomness generation rate. With the experiment settings, $3.125 \times 10^{8}$ quantum states are prepared and measured per second.

\begin{figure*}[htbp]
\centering
\includegraphics[width=15 cm]{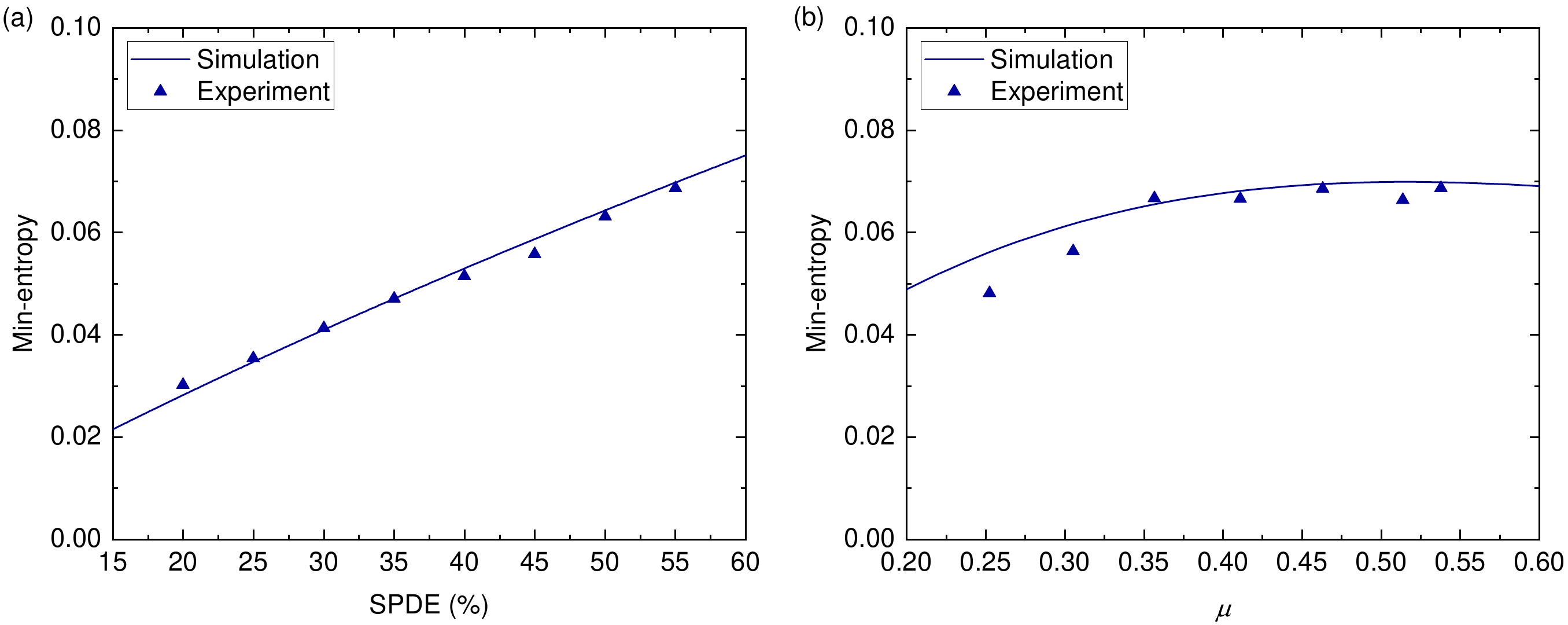}
\caption{Simulation (solid line) and experiment (triangles) results of min-entropy as a function of SPDE (a) with optimized values of $\mu$ and $\nu$ and $\mu$ (b) with 55\% SPDE of an optimized value of $\nu$.
Here the state preparation error rate is set to 3\% according to the measured result shown in Fig.~\ref{figS3}.}
\label{fig4}%
\end{figure*}

Before performing the MDI-QRNG experiment, we verify the system performance by regulating the single-photon detection efficiency (SPDE) of SPAD, corresponding to handling the channel loss.
As shown in Fig.~\ref{fig4} (a), both the simulation and experimental min-entropy with optimal $\mu$ and $\nu$ are significantly improved with increased SPDE.
With a relatively high SPDE setting of $55\%$, we further verify the system performance with different settings of $\mu$ and corresponding optimized settings of $\nu$.
In Fig.~\ref{fig4}(b), although the simulation result of min-entropy always increases with the increase of $\mu$, the experimental min-entropy is close to stable when $\mu$ exceeds $\sim$ 0.35.
The tomography results exhibit excellent stability with different settings, which indicates that the protocol could be robust.
To maximize the randomness generation rate the optimized intensities of $\mu$ and $\nu$ are set to $0.45$ and $0.33$, and the proportion of the tomography mode is set to $1\%$.

\begin{figure}[htbp]
\centering
\includegraphics[width=10 cm]{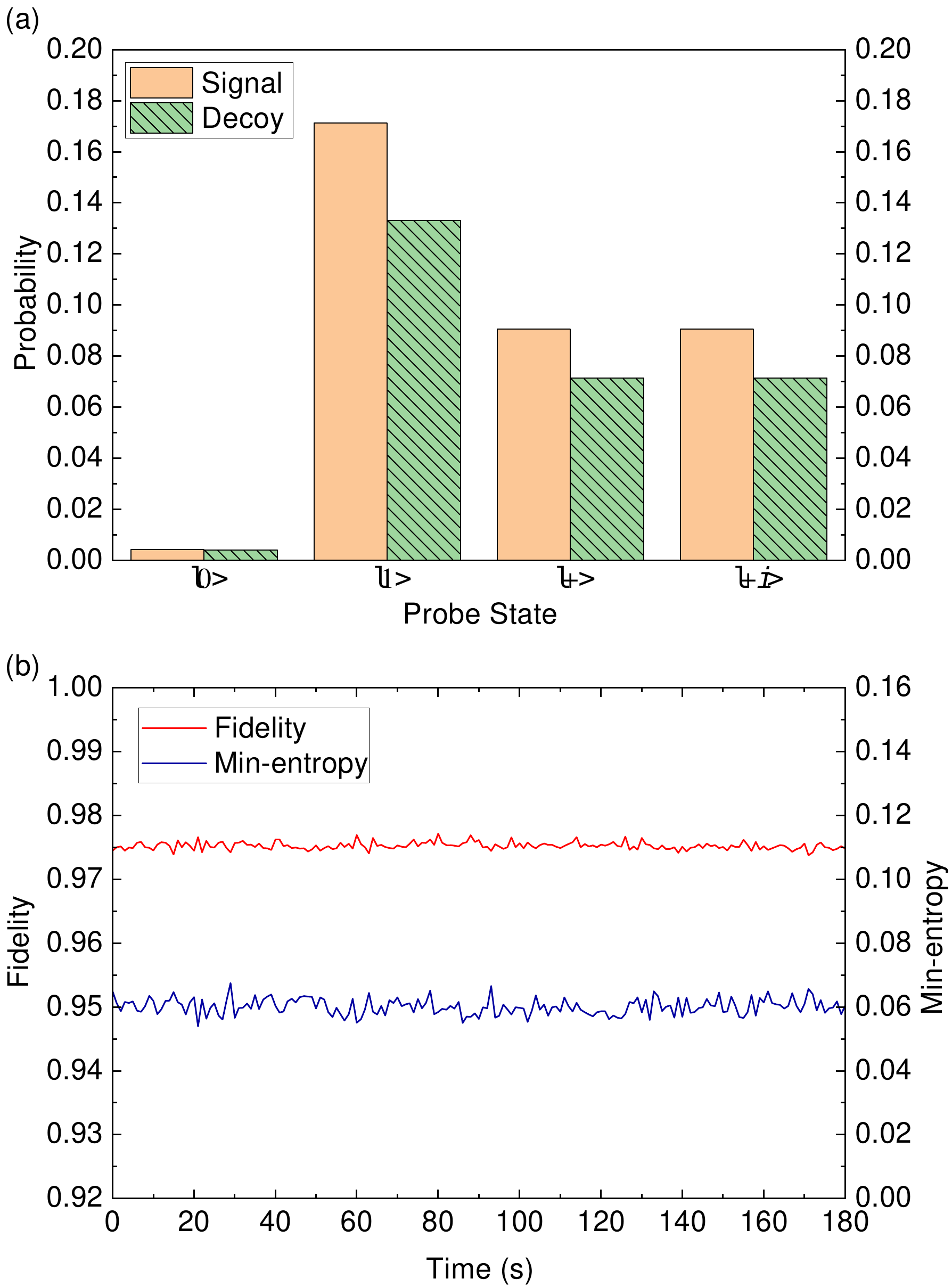}
\caption{(a) Results of the tomography mode in the MDI-QRNG implementation with $5.625 \times 10^{8}$ probe states prepared in $180$ s. (b) Calculated results of the fidelity and min-entropy for each second during $180$ s.}
\label{fig5}%
\end{figure}

During the MDI-QRNG experiment, $5.625 \times 10^{10}$ quantum states are prepared in $180$ s and sent to the measurement device, in which $5.625 \times 10^{8}$ states are randomly designed as probes for the measurement tomography. We set $\eta_j =0.25$ and $p_s=0.5$. The probabilities of $p_\mu(1|j)$ and $p_\nu(1|j)$ defined in Eq.~\eqref{eq:pnu} are illustrated in Fig~\ref{fig5}(a).
Further, by applying the randomness quantification method in Eq.~\eqref{eq:minentropy}, we calculate the lower bound of min-entropy as $7.37 \times 10^{-2}$ bits per pulse, which means that the corresponding quantum randomness generation rate could reach $23$ Mbps.
Here, we neglect the randomness consumption of preparing probe states. By diminishing the proportion of the tomography mode, the overall randomness consumption can be significantly reduced. For instance, lowering the proportion from $1\%$ to $10^{-6}$ results in an overall randomness consumption of $(3+log_2{10^6}) \times 312.5\approx 7.2$ kbps, which can be neglected compared to the generation rate of 23 Mbps.

\subsection{Discussion}
Due to the advantages of the system's high repetition frequency and the SPAD's high count rate, enough probe states can be accumulated for the tomography mode within only one second.
Fig.~\ref{fig5}(b) plots the lower bounds of the fidelity and the min-entropy calculated from the tomography results for each second during the experiment time of $180$ s, from which one can observe the stability of the MDI-QRNG system with a high randomness generation rate. For a brief comparison, we list the generation rates during the previous implementations of MDI-QRNG and Semi-DI QRNG based on single-photon detection in Table~\ref{tab1}.
Specifically, compared with previous MDI-QRNG implementation \cite{Nie16}, the randomness generation rate has been increased by three orders of magnitude.

\begin{table}[htbp]
\centering
\setlength\tabcolsep{10pt}
\caption{Comparison between present work and reported single-photon detection (SPD) based Semi-DI QRNG implementations. MDI: measurement-device-independent, SI: source-independent, Semi-DI: semi-device-independent, UCD: up-conversion single-photon detector, SNSPD: superconducting nanowire single-photon detector, N/A: unavailable.}
\begin{tabular}{ccccc}
    \hline
    Year    & Type      & Rate per pulse  & SPD type &Generation rate\\
    \hline
    2015    & Semi-DI   & N/A                   & InGaAs/InP SPAD& 23 bps    \cite{Brunner15}               \\
    2016    & MDI       & $2.3 \times 10^{-4}$  & InGaAs/InP SPAD& 5.7 kbps  \cite{Nie16}                   \\
    2016    & SI        & $5 \times 10^{-3}$    & Si SPAD        & 5 kbps    \cite{Ma16}                    \\
    2017    & Semi-DI   & $0.33$                & Si SPAD        & 16.5 Mbps \cite{PhysRevApplied.7.054018} \\
    2019    & SI        & N/A                   & UCD            & 1.81 Mbps \cite{li2019QRNG}              \\
    2019    & Semi-DI   & $6.4 \times 10^{-3}$  & InGaAs/InP SPAD& 320 kbps  \cite{Nie19}                   \\
    2019    & Semi-DI   & $0.1$                 & Si SPAD        & 1.25 Mbps \cite{PhysRevA.100.062338}     \\
    2021    & Semi-DI   & $1.36 \times 10^{-4}$ & SNSPD          & 68 kbps   \cite{zhang2021simple}         \\
    2021    & Semi-DI   & $0.23$                & SNSPD          & ~0.76 Mbps\cite{Tebyanian_2021}          \\
    2022    & Semi-DI   & $0.025 \pm 0.005$     & Si SPAD        & 4.4 kbps  \cite{PhysRevApplied.17.034011}\\
    2022    & SI        & $~0.067$              & InGaAs/InP SPAD& 1.34 Mbps \cite{PhysRevLett.129.050506}  \\
    2022    & SI        & $~0.17$               & InGaAs/InP SPAD& 3.37 Mbps \cite{Lin:22}                  \\
    2023    & SI        & $0.1$                 & SNSPD          & 505 kbps  \cite{Liu:23}                  \\
    2023    & SI        & $~0.16$               & InGaAs/InP SPAD& 7.94 Mbps \cite{du2023source}            \\
    Present & MDI       & $7.37 \times 10^{-2}$ & InGaAs/InP SPAD& 23 Mbps                                  \\
    \hline
\end{tabular}
\label{tab1}
\end{table}

\section{Conclusion}
In summary, this work presents a high-speed MDI-QRNG protocol that leverages a robust quantum measurement tomography approach to counteract imperfections in single-photon sources. The experimental demonstration, employing a high-speed time-bin encoding system, results in a record quantum randomness generation rate of up to $23$ Mbps. The high generation rate, combined with the all-fiber implementation, underscores the protocol's potential for practical applications.

Additionally, the generation rate could be further enhanced by using a higher system clock rate and more efficient detectors. For instance, with the integration of multipixel SNSPDs boasting an efficiency of 78\% and a high clock rate of 2.5 GHz \cite{Zhang2019SNSPD, Li2023HighrateQK}, the calculated min-entropy is projected to increase to about 0.129 bits per pulse, potentially pushing the generation rate above 300 Mbps. Moreover, while this work primarily addresses imperfections from multi-photon components, the mispreparation of probe states, particularly in the $X$-$Y$ plane, remains a critical aspect for further investigation.

Finally, we believe that apart from the MDI-QRNG, the measurement tomography could provide a fundamental approach to effectively calibrate the measurement settings for various quantum information processing tasks.

\section*{Acknowledgments}
The authors acknowledge the technical support from W.~Li and the staff of QuantumCTek Co., Ltd. This work has been supported by the National Key R\&D Program of China (2020YFA0309704), the National Natural Science Foundation of China (62275239, 62175227, 12204489, 12174216), the Innovation Program for Quantum Science and Technology (2021ZD0300804), and the Key R\&D Program of Anhui (202203a13010002).

\section*{Data availability statement}
The data that support the findings of this study are available upon reasonable request from the authors.

\appendix
\section{Details on measurement tomography}\label{app:fidelityopt}
We consider the following optimization problem,
\begin{equation}\label{eq:appFopt}
\begin{aligned}
& \min_{a_1,n_x,n_y,n_z} F(\{\Lambda_b^{\mathrm{tom}}\}_b, \{\Lambda_n^{\mathrm{sim}}\}_b) \\
\mathrm{s.t.} \quad & p_1^L(b|j)\leq \mathrm{tr}(\rho_j \Lambda_b^\mathrm{tom}) \leq p_1^U(b|j), \quad b\in\{0,1\},
\end{aligned}
\end{equation}
where $\Lambda_b^{\mathrm{sim}}$ $(b\in \{0,1\})$ is the estimated POVM element by assuming a physical model. In our implementation, the imperfect detector with efficiency $\eta$ can be regarded as a combination of a lossy channel with transmittance $\eta$ and a perfect detector. Then, the actual intensity in each time bin before detection is $\mu\eta/2$. We can calculate the probability of no-click $p_0$, single-click $p_1$ and double-click $p_2$ events as
\begin{equation}
\begin{aligned}
p_0 & = e^{-\mu\eta} \\
p_1 & = 2e^{-\frac{\mu\eta}{2}}(1- e^{-\frac{\mu\eta}{2}}) \\
p_2 & = (1- e^{-\frac{\mu\eta}{2}})^2.
\end{aligned}
\end{equation}
Here, we neglect the effect of dark count for simplicity. Recalling that the non-detection and double click events are recorded to be $'0'$, the estimated POVM is given by
\begin{equation}
\begin{aligned}
\Lambda_0^{\mathrm{sim}} & = p_1 \ket{0}\bra{0} + (p_0 +p_2) I \\
\Lambda_1^{\mathrm{sim}} & = p_1 \ket{1}\bra{1} = I - \Lambda_0^{\mathrm{sim}}.
\end{aligned}
\end{equation}
We then calculate Eq.~\eqref{eq:appFopt} with the optimal $\mu$ and $\nu$. To show how the decoy-state method can improve the tomography accuracy, we also consider the non-decoy case, where only signal states are prepared in the case of conventional measurement tomography. We only need to update the constraints in Eq.~\eqref{eq:appFopt}, i.e., replace the bounds $p_1^L(1|j)$ and $p_1^U(1|j)$ with non-decoy ones. The new bounds $p_1^{L,\mathrm{ndcy}}(1|j)$ and $p_1^{U,\mathrm{ndcy}}(1|j)$ are given by
\begin{equation}\label{eq:nondecoybounds}
\begin{aligned}
&p_1^{L,\mathrm{ndcy}}(1|j)=\max\left\{0,\frac{p^{L,\mathrm{ndcy}}_{\mu}(1|j)-e^{-\mu}p_d-(1-e^{-\mu}-\mu e^{-\mu})}{\mu e^{-\mu}}\right\}  \\
&p_1^{U,\mathrm{ndcy}}(1|j)=\min\left\{1,\frac{p^{U,\mathrm{ndcy}}_{\mu}(1|j)}{\mu e^{-\mu}}\right\},
\end{aligned}
\end{equation}
where
\begin{equation}\label{eq:sfndcy}
\begin{aligned}
  p^{U,\mathrm{ndcy}}_{\mu}(1|j)&=\frac{\mathbb{E}^U(M_{\mu,j})}{N\eta_j} \\
  p^{L,\mathrm{ndcy}}_{\mu}(1|j)&=\frac{\mathbb{E}^L(M_{\mu,j})}{N\eta_j}. \\
\end{aligned}
\end{equation}
\section{Characterization of the InGaAs/InP SPAD}

\begin{figure*}[htbp]
\centering
\includegraphics[width=13.5 cm]{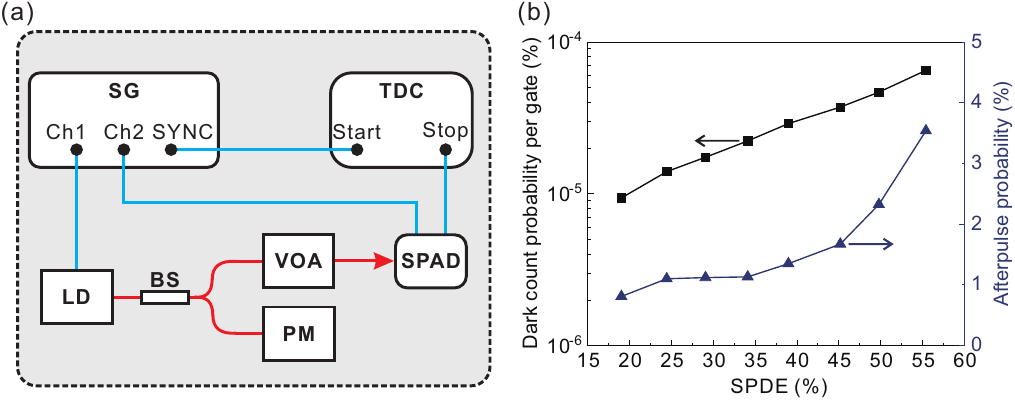}
\caption{(a) Experimental setup for SPAD characterization. (b) Dark count probability per gate and afterpulse probability versus SPDE of the SPAD.}
\label{figS1}%
\end{figure*}

In the experiment, the probe states are measured by a 1.25 GHz sine wave gating InGaAs/InP single-photon avalanche diode (SPAD) \cite{Fang2020}, whose parameters have been characterized including the single-photon detection efficiency (SPDE), the dark count probability per gate $p_d$, the after-pulse probability, and the saturated count rate.
Figure~\ref{figS1}(a) illustrates an experimental setup that follows the single-photon calibration scheme \cite{ZIZ15, Fang2020} to characterize the InGaAs/InP SPADs. One channel (Ch1) of the signal generator (SG) generates a clock of 625 kHz to drive a picosecond pulsed laser diode (LD).
The pulses pass through a 99:1 beam splitter (BS) in which the 99\% port is monitored by a power meter (PM), and the 1\% port is connected with a variable optical attenuator (VOA). The intensity of laser pulses is attenuated down to the single-photon level.
The SG's second channel (Ch2) generates a 10 MHz clock, which is used as the external reference signal of the 1.25 GHz sine wave-gating SPAD system. The synchronized signal (SYNC) of Ch1 and the detection signal are fed into a time-to-digital converter (TDC) as “start” and “stop,” respectively. The TDC performs a timing tag for each detection event and transmits data to a computer for further processing.
With such settings, the primary parameters can be calculated. Figure~\ref{figS1}(b) plots the measured dark count probability per gate and the after-pulse probability with different SPDEs ranging from 20\% to 55\%, which is realized by adjusting the bias voltage of the SPAD.
To achieve a higher saturated count rate and reduce the after-pulse effect of the SPAD, the dead time and the working temperature are set to 4 ns and 303 K, respectively.

\section{Verification of the probe states}

\begin{figure}[htbp]
\centering
\includegraphics[width=10.5 cm]{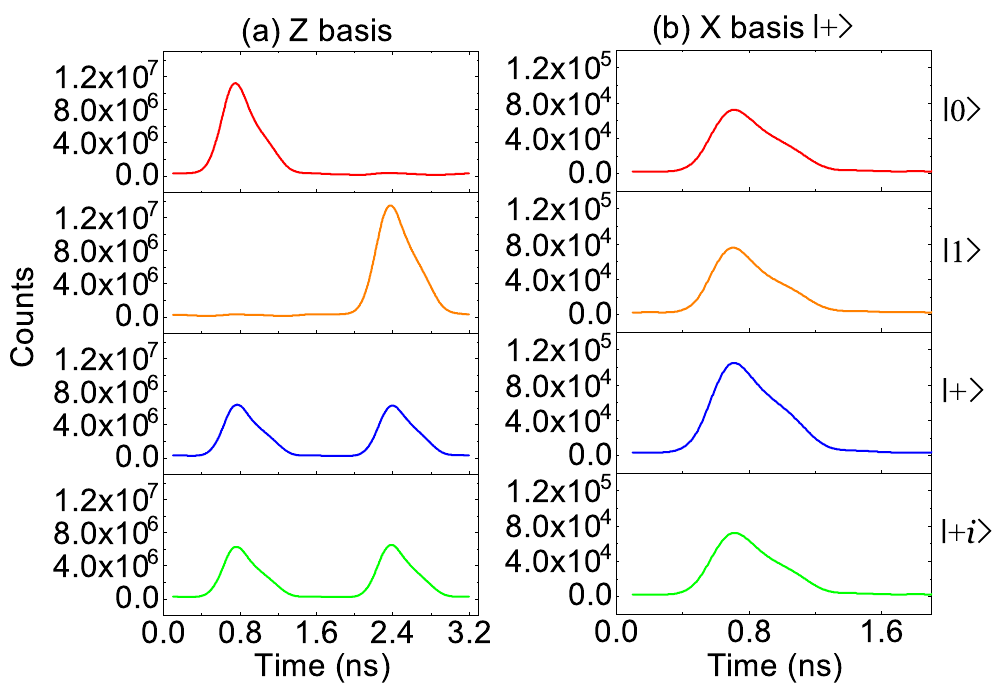}
\caption{Count rate distributions of the prepared quantum states measured in $Z$ (a) and $X$ (b) bases using SPAD and TDC.}
\label{figS2}%
\end{figure}

\begin{figure}[htbp]
\begin{minipage}[b]{1\linewidth}
\centering
\includegraphics[width=8.5 cm]{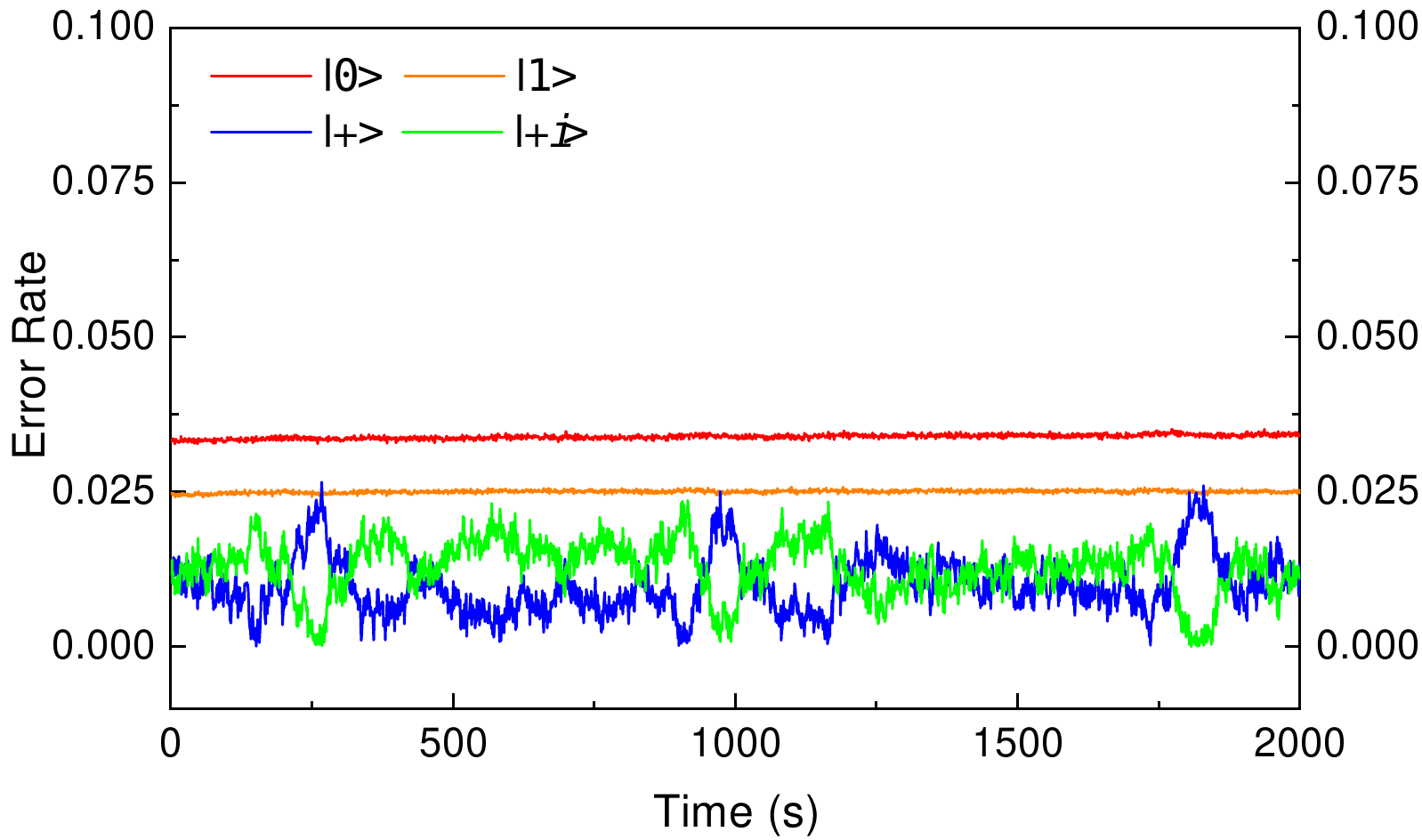}
\end{minipage}
~\\
\begin{minipage}[b]{1\linewidth}
\centering
\centering
\tabcolsep0.1in
\begin{tabular}{c|c}
  \hline
  Prepared state   &{Average error rate}   \\
  \hline
        $\ket{0}$               & $3.39\%$  \\
        $\ket{1}$               & $2.50\%$  \\
        $\ket{+}$               & $0.99\%$  \\
        $\ket{+i}$              & $1.27\%$  \\
  \hline
\end{tabular}
\end{minipage}
\caption{The fluctuations over 2000 s (upper figure) and average values (lower table) of the error rate of the prepared quantum states in the $Z$ basis.}
\label{figS3}
\end{figure}

In the experiment, the required probe quantum states $\ket{0}$, $\ket{1}$, $\ket{+}$, and $\ket{+i}$ are prepared and verified carefully.
As shown in Fig.~\ref{figS2}(a), typical count rate traces of the four time-bin states are directly measured in $Z$ basis, which, however, cannot distinguish $\ket{+}$, and $\ket{+i}$.
To further distinguish $\ket{+}$ and $\ket{+i}$, the $Z$ basis measurement is adjusted to $X$($Y$) basis measurement by inserting a Michelson interferometer with the same delay of the MZI in the source part before the SPAD.
Typical count rate distributions on $X$ basis measurement are shown in Fig.~\ref{figS2}(b).
For clarity, only one output pulse, which is the interference of early and late pulses at the same input time-bin state, is plotted to infer the measured quantum states.
The error rate of the prepared probe states is measured in the preparation stage.
As shown in Fig.~\ref{figS3}, the fluctuations of the error rate over 2000 s and the average error rate values are shown in the upper figure and the lower table, respectively.
Due to the afterpulse effect of SPAD \cite{ZIZ15}, the error rates of states $\ket{0}$ and $\ket{1}$ are higher than those of states $\ket{+}$ and $\ket{+i}$.

\section{Randomness test result}

\begin{figure}[htbp]
\centering
\includegraphics[width=8.5 cm]{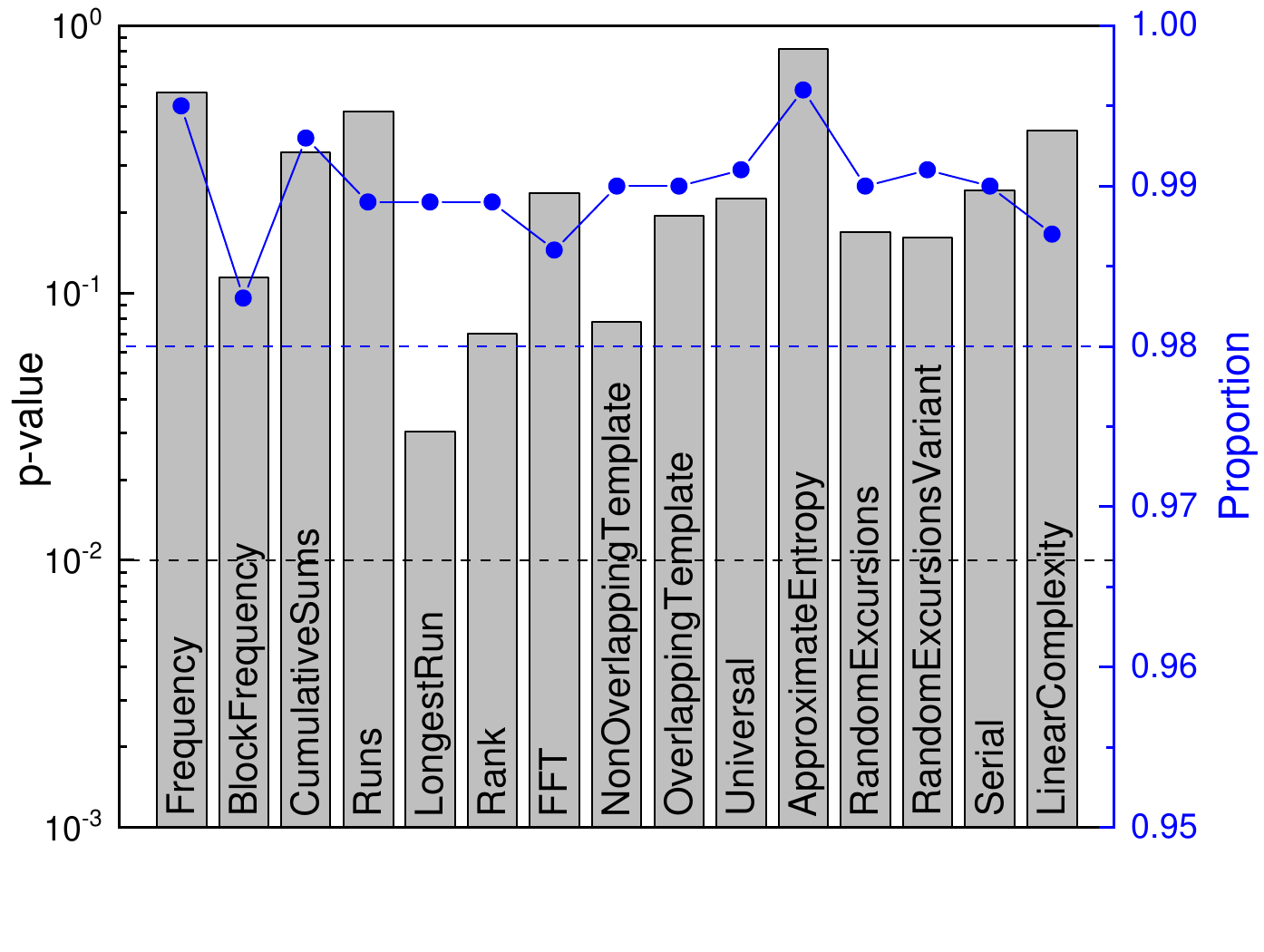}
\caption{Typical NIST test results of the final random data with a size of 1 Gbits. Given a test item, when the p-value (column) and the proportion (dot) are more than 0.01 and 0.98, respectively, the random data pass the item.}
\label{figS4}%
\end{figure}

During the experiment, $5.625 \times 10^{10}$ raw random bits are generated in $180$ s with a min-entropy of $7.37\times 10^{-2}$ bits per bit.
A Toeplitz-matrix hash function is applied for randomness extraction \cite{Ma13Postprocessing}, and more than 4 Gbits of final random numbers are extracted.
Finally, standard NIST statistical tests \cite{NIST} are applied to verify the quality of the final random bits. A typical randomness test result of 1 Gbits final random data is listed in Fig.~\ref{figS4}.

\section*{References}

\bibliographystyle{iopart-num}

\bibliography{bibMT}

\end{document}